\documentclass[aps,prd,showpacs,twocolumn,floatfix,superscriptaddress,preprintnumbers]{revtex4}
\usepackage{amsmath}
\usepackage{amssymb}
\usepackage{epsfig}
\usepackage{graphicx}
\usepackage{stmaryrd}
\newcommand{\lta}{\;
  \raise0.3ex\hbox{$<$\kern-0.75em\raise-1.1ex\hbox{$\sim$
  }}\;\hskip-2pt }
\newcommand{\gta}{\;
  \raise0.3ex\hbox{$>$\kern-0.75em\raise-1.1ex\hbox{$\sim$
  }}\;\hskip-2pt }

\begin{document}

\title{Resonances for activity waves in spherical mean field dynamos}
\author{D. Moss} \email{moss@maths.manchester.ac.uk} 
\affiliation{School of Mathematics, University of Manchester, Manchester M13 9PL, UK}
\author{D.D. Sokoloff$^{\,}$} \email{sokoloff.dd@gmail.com}
\affiliation{Department of Physics, Moscow State University, 119999,
Moscow, Russia}

\date{\today}

\begin{abstract}

We study activity waves of the kind that determine cyclic magnetic activity of 
various stars, including the Sun, as a more general physical rather than
a purely astronomical problem. We try to identify resonances which are 
expected to occur when a mean-field dynamo excites waves of 
quasi-stationary magnetic field in two distinct spherical layers. 
We isolate some features that can be associated with resonances in 
the profiles of energy or frequency plotted versus a dynamo governing 
parameter. Rather unexpectedly however the resonances in spherical dynamos 
take a much less spectacular form than resonances in many more familiar 
branches of physics. In particular, we find that the magnitudes of resonant phenomena 
are much smaller than seem detectable by astronomical observations,
 and plausibly any related effects in laboratory dynamo experiments 
(which of course are not in gravitating spheres!) are also small. 
We discuss specific features 
relevant to resonant phenomena in spherical dynamos,  and find parametric resonance 
to be the most pronounced type of resonance phenomena. Resonance conditions
for these dynamo wave resonances are rather different from those for 
more conventional branches of physics. We suggest that the 
relative insignificance of the phenomenon in this case is because the phenomena of 
excitation and propagation of the activity waves are not well-separated from 
each other and this, together with the nonlinear nature of more-or-less 
realistic dynamos, suppress the resonances and makes them much less pronounced 
than resonant effects, for example in optics,. 
\end{abstract}
\maketitle

\section{Introduction}

The well known solar activity cycle is associated with propagation of a wave of quasistationary magnetic field somewhere within the solar convective envelope. This wave is believed to be excited and supported by electromagnetic induction 
effects driven by the joint action of differential rotation and mirror-asymmetric flows.
This is known as dynamo action, e.g. \cite{S04}. Correspondingly, 
the wave is known as a dynamo wave. More specifically, there are at 
least two, more or less independent, dynamo waves, one propagating 
equatorwards in each solar hemisphere. 

For quite a long time, solar activity waves provided the unique example of dynamo wave propagation, and so the 
phenomenon was not investigated in a wider context. 
Even such a limited statement of the problem provides 
a quite rich range of instructive physical phenomena. 
From time to time the dynamo engine "misfires", and solar 
Grand Minima, including the Maunder Minimum in the XVIIth - early XVIIIth 
centuries, occur (see for review e.g. \cite{SY03}). At the end of the Maunder Minimum,
the dynamo wave propagated in one solar hemisphere only, and a so-called 
"mixed-parity" magnetic configuration appeared \cite{SNR94}. Analysis of XVIIIth 
century sunspot data \cite{A08} gives a hint \cite{Ietal10} that from time to time 
magnetic field becomes concentrated near to the solar equator, rather than in two propagating belts in middle latitudes, and a 
configuration of quadrupole-like (instead of the usual dipole-like) symmetry 
then appears.

Recently the variety of physical phenomenon associated with dynamo wave propagation has enlarged substantially. The first 
attempts to isolate dynamo wave propagation from stellar activity data were undertaken in \cite{BH07,K10}, and generation of oscillating magnetic field, probably indicating dynamo wave propagation, was 
obtained in the VKS dynamo experiment \cite{Metal07}. This opens new perspectives in 
the topic, in particular the opportunity to study dynamo waves as a physical, 
rather than a purely  astronomical, phenomenon. 

\begin{figure}[b]
\begin{center}
\includegraphics[
width=0.48\textwidth]{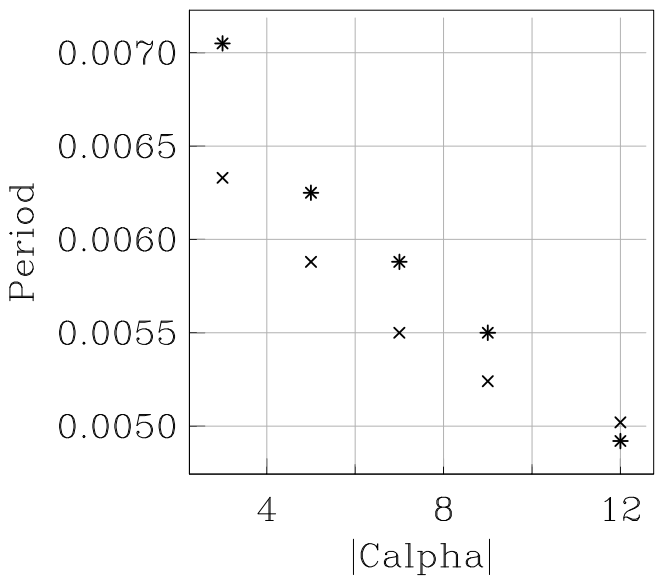}
\caption{Periods of oscillation in energy
when dynamo action occurs in a single layer. 
Crosses:  $C^I_\alpha =0$, $C^{II}_\alpha$ varies; asterisks:  $C^{II}_\alpha =0$, $C^I_\alpha$ varies .}
\label{eberper}
\end{center}
\end{figure}

This change of viewpoint in the investigation of dynamo waves 
has attracted attention to these problems, which had previously remained 
underinvestigated. 
In particular, it was stressed \cite{DG12} that resonance,
which is a prominent and sometimes spectacular phenomenon in 
the conventional theory of various oscillations and wave propagations, must also
play an important role in dynamo wave propagation. We share their view that the topic was not adequately addressed in previous studies, and only a few papers 
(including \cite{KS93,Moss96}) have addressed  the topic. 
The point is that the natural statement of the problem relevant to astronomical
contexts did not focus attention on this issue. 

Here we present our findings associated with the problem of resonance phenomena in dynamo wave propagation. Our general conclusion can be formulated as follows.
Resonances in dynamo systems do occur, and sometimes they can be isolated in 
the solutions of the governing equations (e.g. \cite{KS93,Moss96}). 
However in general dynamo wave propagation as a physical phenomenon differs 
quite substantially from, say, the propagation of acoustic or electromagnetic waves. 
The point is that conventional wave propagation can be easily separated from 
the wave excitation, whereas dynamo wave propagation is usually intimately
associated with dynamo wave excitation. As a result, solutions of
the dynamo equations usually fail to separate in any explicit way 
resonant phenomena from the more general phenomena 
of dynamo wave excitation.

\begin{figure}[b]
\begin{center}
\includegraphics[
width=0.24\textwidth]{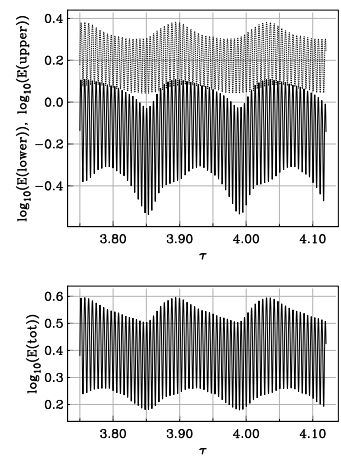}
\caption{Time variations of magnetic energy for $C_\alpha^I=6, C_\alpha^{II}=-5$. Upper panel - toroidal field energy in lower (solid curve) and upper (broken)
layers separately, lower panel - total magnetic energy. }
\label{time}
\end{center}
\label{calphazero}
\end{figure}

We illustrate the problem by numerical investigation of 
a dynamo model in which dynamo waves are excited in two adjacent
spherical shells with different distributions of dynamo governing parameters.
The magnetic field is continuous across the boundary between the two domains.
A related model was discussed previously in \cite{MS07,MSL11} in a more specifically astronomical context. 

\section{The dynamo model}
\label{model}

We consider a standard mean field dynamo equation ($\alpha^2 \omega$-dynamo)

\begin{equation}
{\frac{\partial {\bf B}}{\partial t}} = {\rm curl} \, ({\bf V \times B} + \alpha {\bf B} - \eta \, {\rm curl} \, {\bf B}),
\label{SKR}
\end{equation}
where  $\eta$ is the turbulent diffusivity and $\alpha$ represents the usual isotropic alpha-effect (for details of the model see \cite{MS07};
the codes used there are in turn based on that of \cite{MB00}). 
In Eq.~(\ref{SKR}) $\bf V$ is the velocity from the (differential) 
rotation and we only look for axisymmetric solutions. 
A simple algebraic $\alpha$-quenching (in which $\alpha$ is reduced by a
factor $1+ (B/B_0)^2$,
where $B_0$ is a field strength at which dynamo saturation is expected) is used to suppress dynamo action as the magnetic field grows.

\begin{figure}[b]
\begin{center}
\includegraphics[
width=0.48\textwidth]{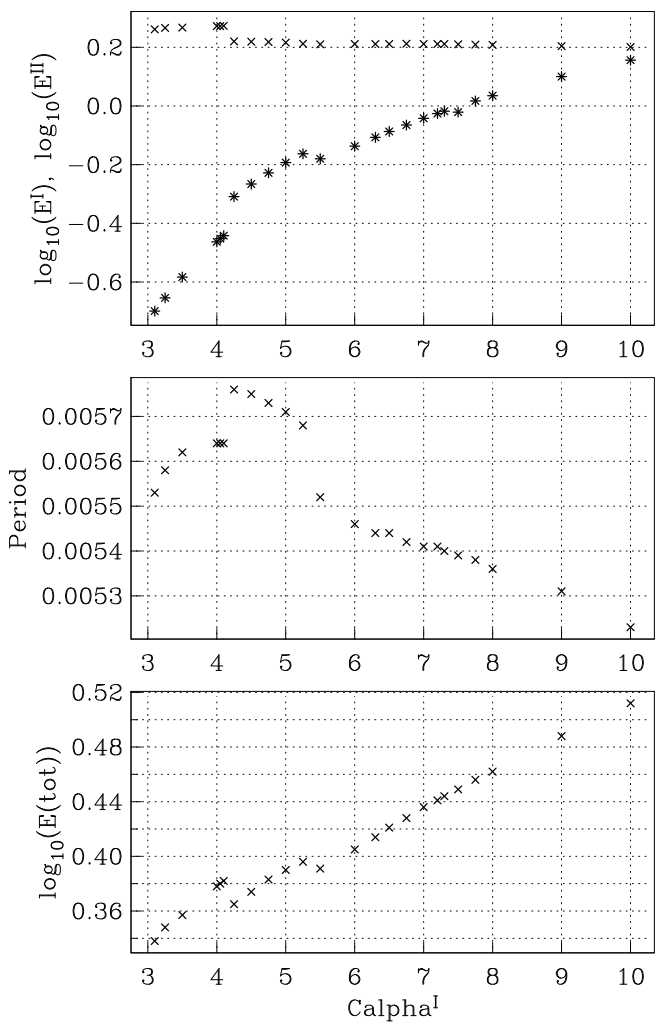}
\caption{The case $C_\alpha^{II}=-5$, $C_\omega^I=C_\omega^{II}=10^5$.
Top panel: toroidal field energies in the lower and upper layers
(asterisks and crosses respectively). The lower panel shows the total energy 
and the middle panel shows the variation of the period of oscillation 
of the energy. In each panel, quantities are plotted as a function of $C_\alpha^I$.}
\label{reson1}
\end{center}
\end{figure} 

We solve the equation in spherical geometry for two adjacent dynamo active 
layers, the 'deep' layer I with $0.7 \le r \le 0.85$ and the 'shallow' layer II with $0.85 \le r \le 1$ ($r$ is the fractional radius). We consider a very simple rotation profile with rotation velocity linear in $r$,
and $\alpha$-profiles in each layer are proportional to $\cos \theta$ where $\theta$ is the polar angle, so that $\theta = 90^\circ$ is the equator. 
$C^I_\alpha$ and $C^{II}_\alpha$ are dimensionless amplitudes of $\alpha$ in 
layers I and II correspondingly, and $C^I_\omega$ and $C^{II}_\omega$ are 
dimensionless amplitudes of the differential rotation. 
We take $C^I_\omega=C^{II}_\omega$ in the examples discussed below.
$\alpha$ is continuous across the interface at $r=0.85$.
All solutions discussed have odd ("dipole-like") parity.

Our investigation is focussed on the more physical aspects of the problem,
but it would be straightforward to rescale the results for particular celestial bodies, or to take a more realistic rotation curve. 

\section{Looking for resonance in dynamo solutions}

\begin{figure}[b]
\begin{center}
\includegraphics[
width=0.4\textwidth]{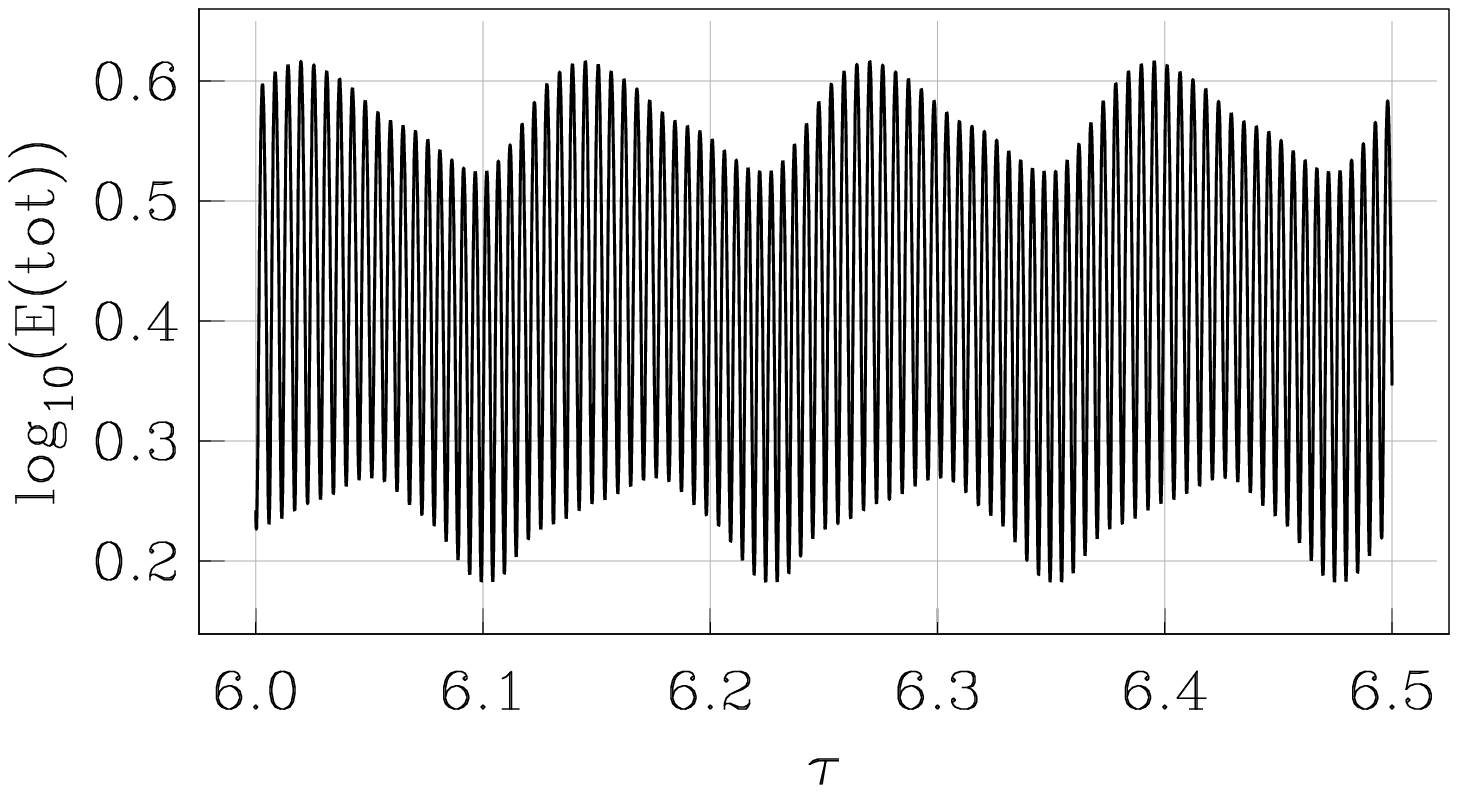} \\
\includegraphics[width=0.4\textwidth]{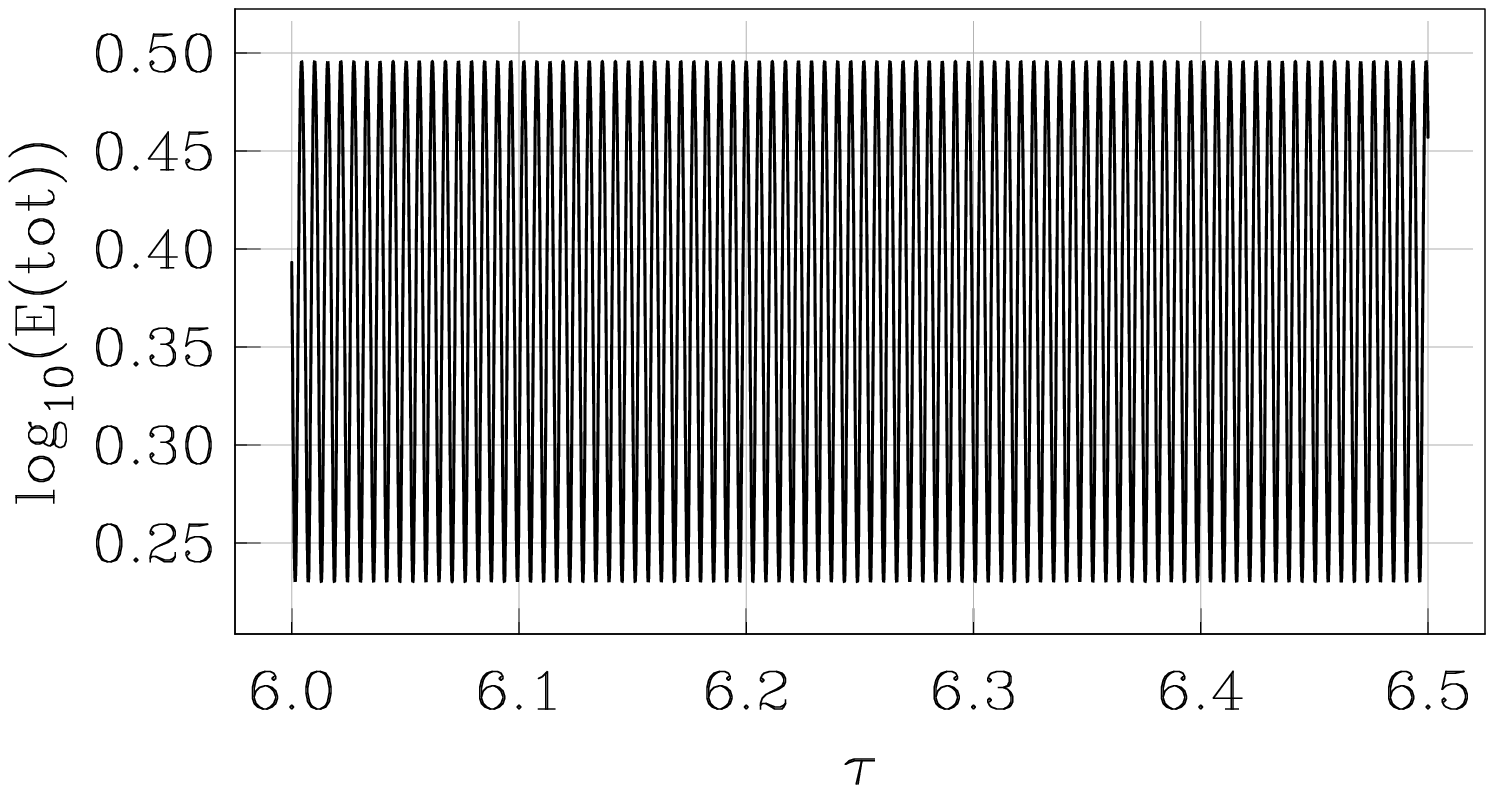}
\caption{Extracts from the time series of total energy for solutions
outside of the resonant domain 
(upper panel, $C^I_\alpha = 6.5$), and inside this domain 
(lower panel, $C^{I}_\alpha = 4.5$). }
\label{four}
\end{center}
\end{figure} 

\subsection{Background}

We obtained solutions for our model for various governing parameters. 
We first determined the periods when dynamo action occurs in one layer
only, i.e. either $C_\alpha^I$ or $C_\alpha^{II}$ is zero.
Note that the magnetic field does penetrate the passive layer,
although its strength is substantially reduced there. 
Here we either take $C_\alpha^I>0$ so that with $C_\alpha^{II}=0$ we obtain
poleward migration, or $C_\alpha^{II}<0$ so that with $C_\alpha^I=0$
there is equatorward migration. The variations of period
with $|C_\alpha^{I,II}|$ for these cases are shown in Fig.~1.

More generally, the model includes two interacting dynamo regions,
which are physically separate but adjacent, which possess specific
individual eigenfrequencies, so quite naturally the resulting 
time-series for the total magnetic energy and magnetic energy 
for the upper and lower layer can be more complicated than 
just a harmonic oscillation; see Fig.~2
where the result for a typical solution ($C_\alpha^I=6, C_\alpha^{II}=-5$) is shown.

\subsection{Resonant solutions}
\label{resonant}

We then set $C_\alpha^{II}=-5$ and allowed $C_\alpha^I$ to vary,
and plotted the period of the total energy 
(defined by a mean over $O(1000)$ oscillations) 
The variations of toroidal field energies in the upper and lower layers, the
(common) period and the total magnetic energy are shown in Fig.~3 as a 
function of $C_\alpha^I$.
We see  a feature, in the form of a local peak of period, in the
region $C^I_\alpha = 5 \pm 1$.
A related feature appears in the plots of  all three quantities (although it
is maybe less pronounced in the lower panel). 
Examination of Fig.~2
shows that this feature does not correspond
to $C_\alpha^I$ values for which the periods of the layers, taken individually,
are equal. For $C_\alpha^{II}=-5$, periods are approximately equal
when $C_\alpha^I\approx 7$.

%
We show in Fig.~4 short samples of the time series of the total energy
for solutions outside of the resonant domain ($C_\alpha^I=6.5$, upper panel),
and inside the domaon ($C_\alpha^I=4.5$, lower panel).

\subsection{Resonance with a subcritical dynamo layer}
\label{subcrit}

\begin{figure}[b]
\begin{center}
\includegraphics[
  width=0.48\textwidth]{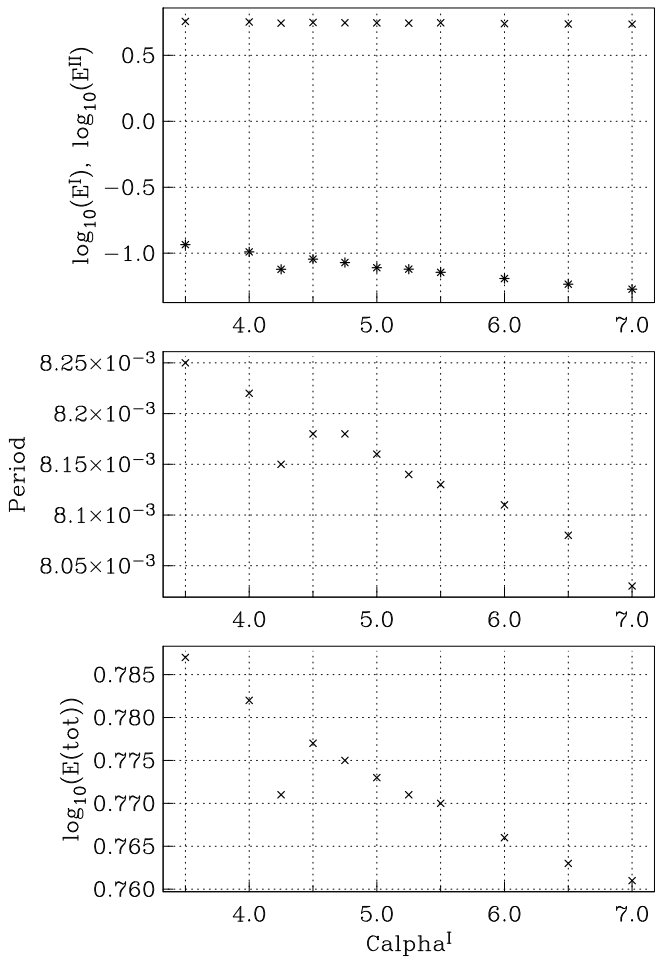}
\caption{The case $C_\alpha^I=-5$, $C_\omega^I=C_\omega^{II}=10^5$,
with thin lower layer, $0.7\le r \le 0.775$.
Top panel: toroidal field energies in the lower and upper layers
(asterisks and crosses respectively). The lower panel shows the total energy 
and the middle panel shows the variation of period.}
\label{reson2}
\end{center}
\end{figure}

We recognize that the resonant behaviour was not very well pronounced in 
the previous results and thus  try to explore and explain this
situation. 
We do not doubt that resonance in its purely mathematical sense occurs 
in dynamo solutions. More specifically, if the 
governing equations for a kinematic dynamo problem have a degenerate 
eigenvalue $\Gamma_n = \gamma_n + i \omega_n$ ($\gamma_n >0$), then 
it is more than plausible that the corresponding set of eigenvectors 
becomes insufficient to provide a basis, and the 
desired solution, apart from terms proportional to $\exp (\gamma_n + i \omega_n) t$, may contain resonant terms 
proportional to  $t \exp (\gamma_n + i \omega_n) t$. 
The point however is that the exponential growth of the solution as 
$\exp \gamma_n t$ or $t \exp \gamma_n t$ occurs just at the initial 
stage of the dynamo wave excitation and is a matter 
of more or less purely theoretical interest. 
A more practical issue is the amplitude of the steady-state oscillation 
obtained after the exponential growth is somehow saturated  -- here by the
term $1+(B/B_0)^2$. 
The saturation reduces the growth rate $\gamma_n$ to zero, 
and this is a much stronger effect than the additional power-low growth 
due to the resonance. The exceptional situation discussed by, say,  \cite{KS93,Moss96}) addresses the cases where excitation of a 
particular solution without a resonance does not occur,
 and it is  the existence of a resonance that allows this solution to develop.
In contrast, resonances in acoustic or electromagnetic wave propagation occur 
when a wave can marginally propagate and does not grow ($\gamma =0$),
 and the resonance is clearly recognisable as a growing (and then saturated) 
solution.

In another investigation, the lower layer was made much thinner 
$(0.7\le r \le 0.775)$, so a dynamo in this layer alone is subcritical (i.e. with $C_\alpha^I=\lta 10, C_\alpha^{II}=0$, dynamo
action does not occur)
and the upper layer  correspondingly thicker, $0.775\le r \le 1.0$.
We keep $C_\alpha^{II}=-5$, $C_\omega^I=C_\omega^{II}=10^5$. There is again
a (smaller) feature attributable to a resonance near $C_\alpha^{I}=4.25$. 
See Fig.~5.

We conclude from Fig.~5
that the resonance here is of a similar nature to that in the previous case.

\subsection{Parametric resonance}
\label{param}

Now we seek the other kind of resonance, i.e. a parametric resonance 
in the dynamo wave driven by
harmonic variations of the dynamo governing parameters. Our preliminary expectations here are twofold. From one hand, 
the resonances recognized in \cite{KS93, Moss96}
for galactic dynamos are parametric.
On the other hand, the parametric resonance 
suggested in \cite{MPS02} to explain some details in the magnetic 
activity of stellar binary systems does not seem very spectacular. 

We imposed a modulation $\alpha=\alpha_0(1+f \cos\omega_P t)$
in the region I, with $500\le \omega_P \le 1500$, $f=0.2$,
keeping $C_\alpha^{II}=-5$. We  started from the unperturbed model 
which has $C_\alpha^I=7.5$ (i.e. outside of the resonant region
discussed in Sect.~\ref{resonant}). 
The corresponding period is $P(E)=0.00539$, so $\omega(E)=1166$.
Results are shown in Fig.~6, where resonant peaks are clearly recognizable.  
 
The feature at $\omega_P\approx 1150$ is quite expected, as it satisfies
the standard resonance condition $\omega_P \approx \omega(E) = 2 \omega(B)$ 
(where $\omega(B)$ is the frequency of the magnetic field oscillations).
The features near $\omega_p \approx 700$ and $\omega_p \approx 900$ 
have to be associated with rather unusual excitation conditions.  
Note that these resonant solutions do not settle to a completely steady 
(maybe modulated) oscillation,
even when run for about 2000 oscillations. 
In the two anomalous cases the period is not quite stable, 
but the averaging process seems robust, in that averages
over different temporal sub-ranges give the same values. Note 
that the energy in the upper and lower
regions separately (top panel in Fig.~6) is the toroidal field energy only,
whereas the "total energy" includes also the poloidal.
Over long enough
time intervals the average period in region II is slightly
longer than that of the total energy (see middle panel of Fig.~6).

\begin{figure}[b]
\begin{center}
\includegraphics[
  width=0.48\textwidth]{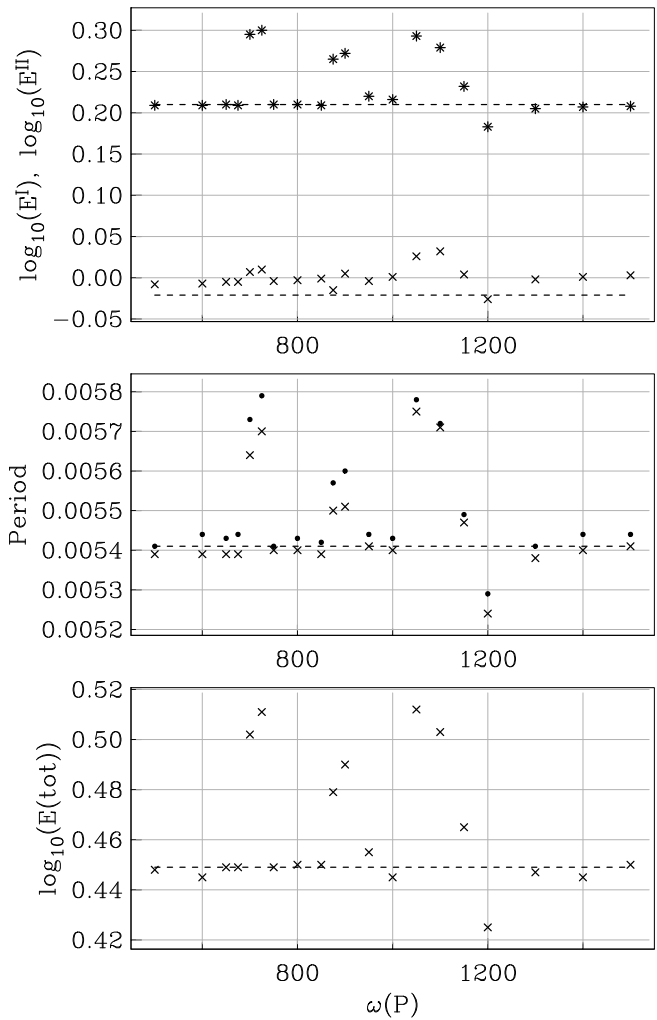}
\caption{ Parametric resonance: notation as in Fig.~3.
 The middle panel now includes the periods of oscillations
 in the upper layer II (shown by small solid circles). 
The dashed horizontal lines
 are values for the unperturbed model. The period of the total energy is plotted by solid dots
 in the middle panel. In each panel, quantities are plotted a functions of $\omega({\rm P})$.
}
\label{reson3}
\end{center}
\end{figure} 

In general, parametric resonance effects seem to be much clearer and 
more pronounced than the resonances analysed in previous models 
(Sects.~\ref{resonant} and \ref{subcrit}). However the peaks in energy profiles 
for parametric resonances are quite broad. 

\section{Discussion}
\label{disc}

The concept of resonance belongs to the basic concepts of 
contemporary physics, and it looks {\it a priori} implausible that it should 
not be applicable to dynamo wave propagation. We have verified this natural 
expectation in the framework of activity waves generated by dynamo waves, 
in a quite traditional model of a 
two-layer mean-field spherical mean-field dynamo of the type conventionally 
considered as a  model for periodic stellar magnetic activity.  

Indeed, we found features in the profiles for energies or periods (frequencies) 
plotted versus a dynamo governing parameter which can be identified 
with a resonance. However, it looks noteworthy that the details resemble more  
discontinuities in the profiles of the behaviour of period and
energy as $C_\alpha^I$ varies, rather than the conventional resonant peaks 
shown in physics textbooks. It is even difficult to claim that the 
resonance makes dynamo excitation more efficient and enhances, say, the
total magnetic energy. The lower panel in Fig.~5 is instructive in this respect: the interaction here appears as a fall in total energy,
 rather than an increase. Of course, resonances are far from being the only 
effects that determine the relevant behaviour in, say, optics, 
and there are many cases where other effects make 
recognition of resonance peaks problematic.  The point 
however is that it is not difficult to 
provide examples of pronounced resonant peaks in optics. 
When presenting our results, we have of course chosen the more pronounced 
examples of resonances for dynamo waves, admittedly from a rather small
sample. 

Correspondingly, we cannot suggest that the sort of resonances discussed here
will leave observable signatures. Our intention is to draw attention to
novel behaviour.
Our results are not so marked as those in \cite{DG12},
arguably because a linear problem was considered there. The inclusion
of a nonlinear saturation mechanism reduces these effects.

The parametric resonances present in the spherical dynamo models investigated 
here (Sect.~\ref{param}) are much clearer than the other resonance features 
(Sects.~\ref{resonant} and \ref{subcrit}). Perhaps, this is a consequence of 
controlling by varying the frequency $\omega_P$, 
which governs the parametric excitation but,
leaves unchanged the "nominal" parameters of the dynamo model for the 
vanishing modulation amplitude $f=0$. 

Our numerical experiment illuminate the unusual excitation conditions 
for parametric resonances. Perhaps, it can be said that the Mathieu equation 
is not a fully adequate model for parametric resonance in spherical dynamos. 
Isolation of the
resonance condition here appears an attractive topic for dynamo theory,
but it is however obviously beyond the scope of this paper. 
The transition time required for the dynamo system to achieve a resonant 
type solution  can be long 
and a qualitative explanation in terms of an appropriate model equation also looks desirable.

We note also the existence of a quasi-resonant phenomenon in a case where
one of the dynamo layers is passive, i.e. with the chosen parameters
the lower layer taken in solution does not support dynamo action.

Perhaps the basic differences between resonant phenomena for dynamo waves 
and those in more conventional branches of physics can be summarized as follows.
It is difficult to separate dynamo wave propagation from dynamo wave excitation, while such 
processes are well separated in many other branches of physics. Correspondingly, eigenvalues of 
the linear (kinematic) dynamo problems take complex values. Coincidence of 
two complex quantities is a more severe
requirement rather than equality of two real valued quantities. 
Dynamo self-excitation is in practice saturated
by some back-reaction of the dynamo generated magnetic field on the 
hydrodynamics. This saturation determines the resulting magnetic energy and 
affects the periods of dynamo waves. The corresponding effects specific to
dynamo action are more important than any potential resonant behaviour. 
In additional, the two layers of dynamo action are separated hydrodynamically 
but magnetic field penetrates from one activity layer to the other, 
even if the dynamo action in one layer remains subcritical. Thus the 
very concept of the frequency of dynamo waves in one activity layer taken alone is not fully applicable. 

In general, it seems that the situation under discussion reflects the every-day wisdom that small problems (here 
resonances) can only spoil your life in the absence of larger ones (in this case other uncertainties associated with 
mean field dynamos are much larger).

\acknowledgements
DS is grateful to financial support from RFBR under grant 12-02-00170-a.

\end{document}